\documentclass[sigconf]{acmart}
\AtBeginDocument{%
  }

\copyrightyear{2025}
\acmYear{2025}
\setcopyright{rightsretained}
\acmConference[IDC '25]{Interaction Design and Children}{June 23--26, 2025}{Reykjavik, Iceland}
\acmBooktitle{Interaction Design and Children (IDC '25), June 23--26, 2025, Reykjavik, Iceland}\acmDOI{10.1145/3713043.3731602}
\acmISBN{979-8-4007-1473-3/2025/06}

\begin{document}

\title[Investigating Youth’s Understanding of Generative Language Models]{Investigating Youth’s Technical and Ethical Understanding of Generative Language Models When Engaging in Construction and Deconstruction Activities}

\author{Luis Morales-Navarro}
\email{luismn@upenn.edu}
\orcid{0000-0002-8777-2374}
\affiliation{%
  \institution{University of Pennsylvania}
  \city{Philadelphia}
  \state{Pennsylvania}
  \country{USA}
}

\renewcommand{\shortauthors}{Morales-Navarro}

\begin{abstract}
  The widespread adoption of generative artificial intelligence/machine learning (AI/ML) technologies has increased the need to support youth in developing AI/ML literacies. However, most work has centered on preparing young people to use these systems, with less attention to how they can participate in designing and evaluating them. This study investigates how engaging young people in the design and auditing of generative language models (GLMs) may foster the development of their understanding of how these systems work from both technical and ethical perspectives. The study takes an in-pieces approach to investigate novices’ conceptions of GLMs. Such an approach supports the analysis of how technical and ethical conceptions evolve and relate to each other. I am currently conducting a series of participatory design workshops with sixteen ninth graders (ages 14–15) in which they will (a) build GLMs from a data-driven perspective that glassboxes how data shapes model performance and (b) audit commercial GLMs by repeatedly and systematically querying them to draw inferences about their behaviors. I will analyze participants' interactions to identify ethical and technical conceptions they may exhibit while designing and auditing GLMs. Then I will investigate the contexts in which these conceptions emerge and how participants' personal interests and prior experiences may relate to their conceptions. I will also conduct clinical interviews and use microgenetic knowledge analysis and ordered network analysis to investigate how participants’ ethical and technical conceptions of GLMs relate to each other and change after the workshop.  The study will contribute (a) evidence of how engaging youth in design and auditing activities may support the development of ethical and technical understanding of GLMs and (b) an inventory of novice design and auditing practices that may support youth’s technical and ethical understanding of GLMs.
\end{abstract}



\maketitle

\section{Introduction}

As generative AI/ML technologies become more widely used, it is crucial to support young people in developing AI/ML literacies \cite{long2020ai, touretzky2023machine}. However, much of the emphasis of efforts to promote AI/ML literacies has centered on preparing young people to use these systems \cite{lee2024ai, kalantzis2025literacy}, with less attention to how they can participate in designing and evaluating them. Child-computer interaction and computing education studies argue that design and evaluation activities can support learners to develop their agency and build deeper understanding of sociotechnical systems through the construction and deconstruction of artifacts \citep{dindler2020computational, oleson2020role}. At the same time, most research conducted on youth’s understanding of generative language models (GLMs) has focused on their everyday interactions as users of systems such as ChatGPT \citep{solyst2024children,marx2024identifying}. These studies have investigated youth’s everyday technical and functional understanding of how models work \citep{marx2024identifying} and youth’s critical and ethical stances towards these systems \citep{solyst2024children,morales2024s}. However, technical and ethical issues of machine learning are closely intertwined \citep{10.1145/3531146.3533158}, and there is a need to explore how learners’ technical and ethical conceptions relate to each other.

This dissertation investigates teenagers understanding of generative language models from the inside-out when engaging in construction by designing GLMs and from the outside-in when engaging in deconstruction by auditing GLMs. In this dissertation, I investigate how engaging youth as designers and auditors of generative language models may support the development of their understanding of how these systems work from both technical and ethical perspectives. 

\section{Related Work}
\subsection{Computational Empowerment and AI/ML}
Research on computational empowerment (CE) centers on construction and deconstruction participatory design activities in which learners are ``protagonists'' \citep{smith2023research}. That is, where learners  are the ``main agents'' in designing and evaluating systems \citep{iversen2017child}. Here the conjecture is that through construction and deconstruction activities, learners develop understandings of how computing systems work and skills to design and reflect on computing technologies and their role in everyday life \citep{iversen2017child}. 

Traditionally, it is assumed that construction involves learning about technical aspects to design computing systems for others, while deconstruction involves critical assessment and reflection of computing systems that already exist in the world \citep{iivari2023computational}. However, as \citet{iivari2023computational} argue, construction also involves critically analyzing design ideas. Bringing together CE and computational literacies \cite{kafai2022revaluation}, I argue that AI/ML construction and deconstruction activities must address the functional, critical, and personal dimensions of computational literacies. Construction and deconstruction should support youth to develop both ethical and technical understanding of AI/ML systems through social interaction with the world and in ways that connect to learners' personal interests and identities. What I mean by this is that construction also involves critical aspects of literacies, as learners make ethical and justice-related decisions about how models are designed. Similarly, deconstruction activities involve functional aspects of literacies, as learners need to develop proficiency and skills to be able to evaluate how AI/ML systems work.

Various efforts in CCI have engaged youth in the construction of AI/ML models. These studies have focused on how young people create and label datasets for image, pose, and movement classification tasks \citep{arastoopour_irgens_characterizing_2022, druga_landscape_2022, hjorth2021naturallanguageprocesing4all, tseng_plushpal_2021, bilstrup2024ml}. Several studies have also investigated how youth build language models to classify words or sentences \cite{hjorth2021naturallanguageprocesing4all, norouzi_lessons_2020, katuka2024integrating, alvarez2022socially, jiang2023high, chao2023exploring}. These studies highlight ways youth can engage with natural language processing, connecting data-driven exploration to real-world applications, but they predominantly focus on classification tasks. In these studies, participants created small data sets that can be easily and quickly refined to train models and improve their performance \citep{zimmermann-niefield_youth_2020, tseng2024co}. \citet{vartiainen2021machine} argue that this approach to ML fosters ``data-driven reasoning and design,'' which entails considering dataset design decisions to explain the behaviors of ML systems. Recently, researchers outlined data practices that youth engage with when building models, including practices related to understanding a task, collecting data, understanding data, preparing data, implementing a solution, evaluating performance, deploying, and monitoring \citep{tseng2024co, olari2024data}. 

Several approaches have been proposed to engage young people in the deconstruction of AI/ML systems. The DORIT framework \citep{dindler2023dorit} was designed to support youth's critical inquiry on everyday computing technologies, that is, the investigation of the implications and limitations of computing systems. RAD \citep{10.1145/3641554.3701966} was created to scaffold critiquing AI/ML systems by supporting youth in recognizing harms, analyzing societal aspects of harmful algorithmic behaviors, and deliberating how AI/ML systems could become less harmful. While DORIT and RAD support youth in reflecting on the implications and limitations of computing systems, these frameworks do not engage learners in systematically evaluating algorithmic behaviors. The auditing in five steps approach \citep{morales2025learning}, developed by our team, was designed as a way to support young people to participate in algorithm auditing. Algorithm auditing is an effective strategy in algorithmic accountability and human-centered computing research to systematically investigate and comprehend how AI/ML systems behave from the outside-in. Auditing is systematic as it involves ``repeatedly querying an algorithm and observing its output in order to draw conclusions about the algorithm's opaque inner workings and possible external impact'' \citep{metaxa2021auditing}. The five steps include scaffolding learners to (1) develop a hypothesis about a potentially harmful algorithmic behavior, (2) generate a set of systematic, thorough, and thoughtful inputs to test the hypothesis, (3) run the test, (4) analyze the data, and (5) report the results. As such, auditing does not only involve critical aspects of literacies (e.g., considering the implications and limitations of ML systems) but also functional aspects related to designing an audit, collecting data and analyzing it. I build on these experiences to investigate how youth engage in the construction and deconstruction of generative language models and how they develop their technical and ethical understanding of GLMs. 

\subsection{An In-Pieces Approach to Studying Conceptions of AI/ML}

In this dissertation, I propose taking an in-pieces approach to study teenagers' understanding of GLMs when engaging in construction and deconstruction activities. I build on research on knowledge-in-pieces \citep{disessa2004coherence}, ideology-in-pieces \citep{philip2011ideology}, and folk algorithmic theories \citep{10.1145/3173574.3173694}. \citet{disessa1993toward} argued that everyday intuitive knowledge could be organized as a collection of pieces of knowledge. These pieces enable people to explain and predict phenomena intuitively and in a way that is consistent with their lived experiences. Yet, these exist in a complex system and are contextually activated; as such, novices may have inconsistent understandings developed across different contexts. Here, learners develop a repertoire of pieces of knowledge, adding understandings from instruction to those from their lived experiences while grappling with multiple and sometimes conflicting conceptions they may have about phenomena \citep{Linn_2005}. Some aspects of KiP are also present in \citet{10.1145/3173574.3173694} conceptualization of folk algorithmic theories, where they argue that these theories have a ``fragmentary nature'' characterized by instability and incoherence where ``pieces of information'' activated together are combined to build folk theories. An in-pieces approach may also be useful for studying and understanding critical and ethical conceptions. \citet{philip2011ideology} synthesized KiP with theories of ideology into what he calls ideology-in-pieces to study sensemaking of issues of power. He claims that ideas of justice, ethics, blame, and responsibility do not have normative meanings but meanings that are constructed by communities. Philip argues that when making sense of critical issues related to race, people rely on existing conceptions or ``naturalized axioms.'' These are socially situated within particular systems of power and are activated in specific contexts. Taking an in-pieces approach, I argue, may support the analysis of how technical and ethical conceptions evolve and relate to each other.

\section{Methods}

I started conducting a series of participatory design workshops with 16 high school youth (9th graders) enrolled in a four-year after-school program at a science center in the Northeastern United States. During the workshops, construction activities in which youth build models take a data-driven approach that glassboxes how data shapes model performance and blackboxes the role of learning algorithms in the ML pipeline \citep{10.1145/3459990.3460712, olari2024data, morales2024unpacking}. These activities support youth to iteratively build small datasets (70-350K tokens) to train models using the nanoGPT framework \citep{KaparthyNano}. Deconstruction activities center around algorithm auditing, which involves the repeated querying of AI/ML systems to draw inferences about their opaque inner workings and possible external impacts from the outside-in \citep{metaxa2021auditing}. This method has previously been adapted to successfully scaffold youth in investigating the functionality and potential limitations of generative models \citep{morales2025learning}. In the workshop, youth will audit outputs generated by ChatGPT. 

Using thematic analysis \cite{braun2012thematic} and explanatory \citep{yin2018case} case studies, I investigate the ethical and technical conceptions that youth exhibit while designing and auditing GLMs. In particular, I am interested in the practices, interactions, and peer and mentor conversations leading up to those in-the-moment conceptions and how participants personal experiences, interests and identities relate to those conceptions. I ask:
\begin{itemize}
    \item What ethical and technical conceptions do youth exhibit in-the-moment as they design and audit GLMs? 
    \item What practices, interactions, and peer and mentor conversations lead up to those in-the-moment conceptions exhibited while designing and auditing GLMs? 
    \item How do participants identities, personal interests, and prior experiences relate to the ethical and technical conceptions they exhibit while designing and auditing GLMs?
\end{itemize}
Through clinical interviews \cite{Sherin2012SomeInterviews}, microgenetic knowledge analysis \cite{disessa:2015}, and ordered network analysis methods \cite{tan2024epistemic}, I investigate how participants’ ethical and technical conceptions of GLMs during clinical interviews change from pre- to post- and address the following questions: 
\begin{itemize}
        \item What are participants' ethical and technical everyday conceptions of GLMs before participating in design and auditing activities?
        \item What are participants' ethical and technical conceptions of language models after designing and auditing GLMs?
        \item  How do the relationships between participants' ethical and technical conceptions change from pre to post?
\end{itemize}

\subsection{Current Work and Next Steps}

This March, I started conducting PD workshops with participants and finished doing pre-interviews with them. Over the past year, I have also conducted pilot studies on young people's everyday understanding of AI/ML systems \cite{morales2024investigating, morales2024s}, on youth as designers of babyGPTs \cite{MoralesNavarro2025baby} and auditors of generative AI/ML systems \citep{morales2024youth, morales2025learning}. 

I will continue conducting PD workshops and collecting data through the summer of 2025. Then I will spend the next nine months analyzing data and writing three papers. 

\subsection{Expected Deliverables and Contribution}

With this work, I expect to author three papers that address how engaging in construction and deconstruction activities may support learners to develop their understanding of GLMs. A first paper will center on how designing GLMs may support participants in developing their understanding of how these systems work from the inside-out. A second paper will center on how auditing GLMs may support participants in developing their understanding of how these systems work from the outside-in. Finally, a third paper is concerned with how pieces of knowledge or conceptions that youth may have change after participating in construction activities in which they design GLMs and deconstruction activities in which they audit GLMs. I hope this work will contribute (a) evidence of how engaging youth in design and auditing activities may support the development of ethical and technical understanding of GLMs and (b) an inventory of novice design and auditing practices that may support youth’s technical and ethical understanding of GLMs.

\begin{acks}
I am grateful to my dissertation committee, Yasmin B. Kafai, Danaé Metaxa, Bodong Chen, and Amy J. Ko, for their support. Pilot work for this study was supported by National Science Foundation grants \#2414590, \#2333469, \#2342438. Any opinions, findings, and conclusions or recommendations expressed in this paper are those of the author and do not necessarily reflect the views of NSF or the University of Pennsylvania. 
\end{acks}


\bibliographystyle{ACM-Reference-Format}
\bibliography{references}

\appendix

\end{document}